\documentclass{ws-ijmpe}
\usepackage{xcolor}
\usepackage[super,compress]{cite}
\usepackage[verbose]{hyperref}
\hypersetup{colorlinks=false,allbordercolors=blue,pdfborderstyle={/S/U/W 1}}

\begin{document}

\markboth{A. V. Anufriev, V. N. Kovalenko}{Application of the holographic equations of state for modeling experiments on heavy ion collisions}

\catchline{}{}{}{}{}

\title{Application of the holographic equations of state for modeling experiments on heavy ion collisions}

\author{A. V. Anufriev}

\address{Saint Petersburg State University. \\
7/9 Universitetskaya Nab., St. Petersburg 199034, Russia\\
anton.anufriev@spbu.ru}

\author{V. N. Kovalenko}

\address{Saint Petersburg State University. \\
7/9 Universitetskaya Nab., St. Petersburg 199034, Russia\\
v.kovalenko@spbu.ru}

\maketitle

\begin{abstract}

In this paper, we propose a method for numerical modeling of the nuclear matter properties within the framework of relativistic heavy-ion collisions using a holographic equation of state. Machine learning methods were applied to address the regression and optimization issues during the calibration of the relevant parameters using the LQCD results for quark masses that approximate the physical values. Numerical simulations are performed using the iEBE-MUSIC and vHLLE-SMASH frameworks, which incorporate certain relativistic hydrodynamics solvers. We modify the code by implementing a tabulated holographic equation of state, enabling simulations of quark-gluon plasma evolution with dynamically generated initial conditions via the 3D Monte Carlo Glauber Model and SMASH. Finally, the spectra of produced hadrons are computed using a hybrid iSS+UrQMD and Hadron Sampler+SMASH approaches at the freeze-out stage.
\end{abstract}


\ccode{PACS Number(s): 25.75.-q, 12.38.Mh, 24.10.Nz}

\section{Introduction}
The study of the quark-gluon plasma (QGP) phase diagram is becoming particularly relevant in modern research. The QGP is a special phase of matter that is assumed to have collective properties similar to a relativistic liquid~\cite{1}. In recent years, the hydrodynamic model, which has been demonstrated to be highly effective in describing multiple particle generation, has gained significant popularity in the study of the QGP evolution~\cite{2}. The following text is intended to provide a comprehensive overview of the subject matter.

The distinctive characteristics of the collective existence of quarks and gluons within this theoretical framework are expressed in the specific thermodynamic behavior of the QGP. This is defined by the equation of state (EoS), the general form of which remains unknown. The calculation of the EoS as a function of temperature is enabled by the consideration of Lattice QCD at baryon potentials close to zero, under the assumption of a smooth crossover in this region~\cite{3}. However, for nonzero values of $\mu_B$, which are generally associated with the presence of a critical point in the quark-gluon plasma (QGP), a ``sign problem'' arises due to the uncertainty of the fermionic determinant~\cite{4}. This is why more exotic variants are also considered.

In 1998, Maldacena proposed a rigorous formulation of AdS/CFT invariance~\cite{5}, which further allowed us to consider the advantages of this approach in the low-energy limit of string theory, opening up the working domain of AdS/QCD duality~\cite{6}. This approach assumes the duality of quantum chromodynamics (QCD) and gravitational theory in anti-de Sitter space.

A notable example of this approach is the cycle of results of the I. Ya. The Arefieva group's approach~\cite{7} involves the introduction of an additional scalar dilaton field within the classical Einstein-Maxwell action, complemented by the implementation of a potential reconstruction method. This method leads to the emergence of free parameters within the deformed initial paragraph of the gravitational theory. The subsequent calculation of the dilaton potential is derived from the equations of motion.

The assumption of zero quark masses, which is characteristic of the chiral limit in lattice quantum chromodynamics (QCD) studies, is effectively restored by selecting a deformation factor with a certain set of predicted free parameters for the initial ansatz, as was done in Ref.~\refcite{8}, for example. The present study will also address the issue of selecting the aforementioned factor and adjusting the corresponding free parameters. Another benefit of the methodologies proposed by Arefieva's group is the observation of anisotropy in the spatial components of the 4-dimensional part of the metric. This anisotropy may be indirectly related to the phenomenon of plasma isotropization. The importance of this holographic hypothesis is confirmed by the study of the dependence of the experimentally observed multiplicity density on the collision energy~\cite{9} .

The results of works devoted to holographic equations of state frequently remain theoretical, rarely used to study available experimental data and make future predictions. This situation is a key motivating factor for the search for the possibility of using holography for numerical simulation of the evolution of QGP in a kinematic region of ion-ion relativistic collisions.

\section{The holographic approach applied}
In this research, EoS of the QGP will be obtained by employing a special type ``bottom-up'' soft-wall (SW) approach. This connects the quasi-conformal QCD theory with classical gravity in the AdS space of dimension 5, as proposed in Ref.~\refcite{8}. An action incorporating simultaneously two dilaton fields of the following form is introduced:
\begin{equation}\label{eq1}
 S=\frac{1}{16\pi G_5}\int d^5x \sqrt{-\det(g_{\mu\nu})} \left[R-\frac{f_1(\phi)}{4}F_{(1)}^2-\frac{f_2(\phi)}{4}F_{(2)}^2-\frac{1}{2}\partial_{\mu}\phi \partial^{\mu}\phi-V(\phi)  \right],   
\end{equation}
where $\phi$ is a scalar dilaton field, its potential of unknown type is $V(\phi)$, $F_{MN}^{(a)}=\partial_MA_N^{(a)}-\partial_NA_M^{(a)}$, $f_i(\phi)$ are functions of interaction with Maxwell fields, $i = 1,2$, 
$\det(g_{\mu\nu})$ is the determinant of the metric, and $G_5$ is the gravitational constant in AdS$^5$.

 A special form of the ansatz is introduced in addition to (\ref{eq1}):
\begin{equation}\label{eq2}
ds^2=\frac{R^2}{z^2}b(z)\left[ -g(z)dt^2+dx^2+\left(\frac{z}{R}\right)^{2-\frac{2}{\nu}}dy_1^2+\left(\frac{z}{R}\right)^{2-\frac{2}{\nu}}dy_2^2+\frac{dz^2}{g(z)} \right],
\end{equation}
where $R$ is a dimensional factor that corresponds to the AdS radius for the Poincare metric (we use $R=1$ for further calculations. This is a typical choice for theoretical works and does not reduce the generality of the above expressions), $g(z)$ is the blackening function that determines the thermodynamic behavior of the black brane. The parameter $\nu$ controls the spatial anisotropy of the metric components, 
where $\nu=1$ 
corresponds to the isotropic case.

The deformation factor $b(z)$ from Eq. (\ref{eq2}) in Ref. ~\refcite{8} corresponds to the ``light quarks'' model and was chosen in such a way as to restore the results of lattice QCD calculations in the limit of $m_q\to 0$. Thus:
\begin{align} \label{eq3}
b(z)=\exp({2A(z)}), \\
A(z)=-a\ln(bz^2+1). \label{eq4}
\end{align}

An alternative form of such a factor with fourth power of logarithmic z-dependence was proposed in recent work~\refcite{10}, which will also be used for our study:

\begin{equation} \label{eq5}
A(z)=d \ln(az^2+1)+d \ln(bz^4+1). 
\end{equation}

The expressions corresponding to the thermodynamics of a black hole were obtained by solving the equations of motion from Eqs. (\ref{eq1}) and (\ref{eq2}) with (\ref{eq3}). Taking into account the unit assignment of the constant $L$, determined by the characteristic radius AdS$^5$ $L=1$, thermodynamic parameters of QGP will take the following form:
 \begin{equation} \label{eq6}
T=\frac{1}{4\pi}\left| -\frac{(1+bz_h^2)^{3a}z_h^{1+\frac{2}{\nu}}}{I_1}\left[ 1-\frac{2\mu^2ce^{cz_h^2}}{(1-e^{cz_h^2})^2}\left(1-e^{-cz_h^2}\frac{I_2}{I_1}\right)I_1 \right]  \right|.
\end{equation}

In this expression:

 \begin{align*}
I_1=\int\limits_0^{z_h}(1+b\xi^2)^{3a}\xi^{1+\frac{2}{\nu}}d\xi. \\
I_2=\int\limits_0^{z_h}e^{c\xi^2}(1+b\xi^2)^{3a}\xi^{1+\frac{2}{\nu}}d\xi
 \end{align*}

The result for the entropy density has the form:
 \begin{equation} \label{eq7}
s=\left(\frac{1}{z_h}\right)^{1+\frac{2}{\nu}}\frac{(1+bz_h^2)^{-3a}}{4G}, 
\end{equation}
where $G$ is a dimensionless gravitational constant.

A characteristic relationship between baryon density and potential, independent of the anisotropy parameter $\nu$, is the following:
 \begin{equation}\label{eq8}
\rho=-\frac{c\mu}{1-e^{cz_h^2}}.
\end{equation}

For the numerical simulations, it is necessary to take into account the dependence of EoS on the baryonic potential. Therefore, the pressure in this model is calculated using the following formula:
 \begin{equation}\label{eq9}
p=-\int\limits_0^{z_0}s\frac{dT}{dz_h}dz_h-\int\limits_0^{\mu_0}s\frac{dT}{d\mu}d\mu+\int\limits_0^{\mu_0}\rho d\mu.
\end{equation}

The energy density can be found from the basic thermodynamic identity for relativistic hydrodynamics, combining the eqs. (\ref{eq3})--(\ref{eq9}):
 \begin{equation}
\varepsilon=-p+Ts+\rho \mu.
\end{equation}

\section{Calibrating the model}

The following ideas, which led to Fig.1 and Tab.1, were initially presented in our previous work~\refcite{10}.

A parameter $c=0.227$~GeV$^2$, which appears in the initial metric form (\ref{eq1}), is adjusted to the Regge spectra of the $\rho$ mesons based on the algorithm described in Ref.~\refcite{11}. The parameters $a$ and $b$ (as well as the parameter $d$ for the alternative deforming factor in formula (\ref{eq5}) are to be determined by comparing the prediction of the model with the lattice calculations. The results of Ref.~\refcite{12} are used because the physical masses of the $\rho$ mesons are taken into account. This choice is related to the physical case of quark masses. 

Italian authors~\cite{13} proposed the method of the holographic model calibration which we adopt for this work. Within this method, dimensionless thermodynamic quantities are considered in the form of (\ref{eq6}--\ref{eq9}), with additional factor - energy scale $L$. It converts model's quantities into GeV with the gravitational horizon $z_h$ has units of GeV$^{-1}$. As a result, the power of GeV corresponding to a given physical quantity is determined by the power of the energy scale. Consequently, the scale $L$ is determined with $a$ and $b$.

The isotropic model is calibrated using the results of calculations of $\dfrac{s}{T^3}$ value, while the anisotropic case is adjusted to match the results of the quark susceptibility $\dfrac{1}{T^2}\dfrac{\partial \rho_q}{\partial \mu_q}$. The parameters were adjusted using the least squares method, corresponding results shown in Table 1 (the values of $b$ and $c$ are made dimensionless using the $L$ scale). 

\begin{table}[pt]
\tbl{Values of parameters obtained from the least squares method for standard deforming factor\label{tab1}}
{\begin{tabular}{@{}cccccc@{}}
\toprule
Model       & $\nu$ & $a$     & $b$     & $G$    & $L,$ GeV     \\ 
\colrule
Isotropic   & 1     & 3.71  & 0.011 & 0.34 & 1.08  \\ 
Anisotropic & 4.5   & 3.949 & 0.034 & 0.81 & 1.01  \\ 
\botrule
\end{tabular}}
\end{table}

The fit results for the isotropic and anisotropic models are shown in Fig. 1.
The results of these fits exhibit lower quality compared to, for example, those in Ref.~\refcite{14}.

\begin{figure}[th]
\centerline{
\includegraphics[scale=0.45]{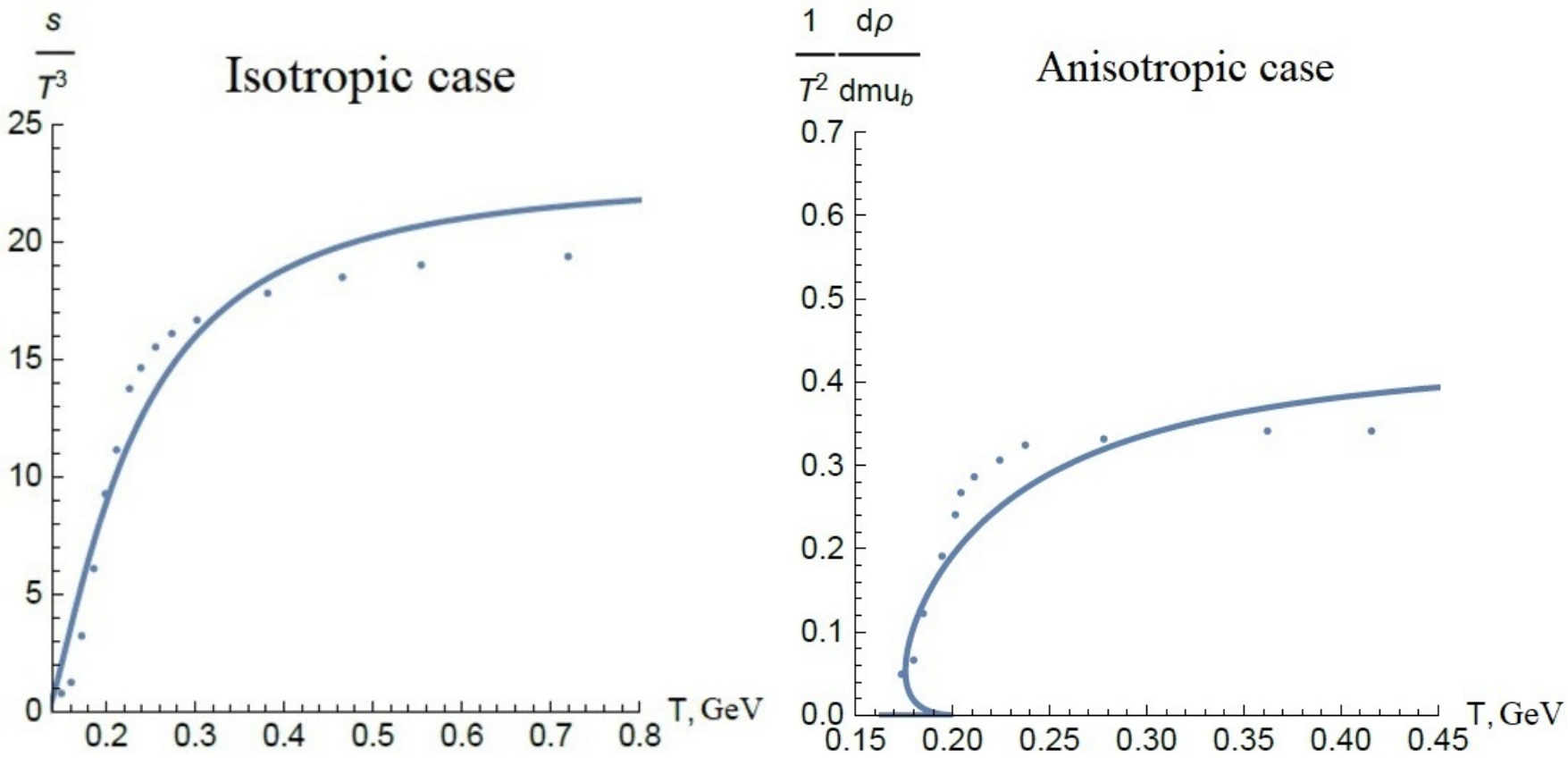}
}
\caption{Result of fitting $\frac{s}{T^3}$ ratio with the data from Ref.~[\citenum{12}] for isotropic (on the left) and anisotropic (on the right) models with standard deforming factor}
\label{f1}
\end{figure}

However, since the model in this case is adjusted via the potential reconstruction method---where the dilaton potential is computed rather than predefined, and all free parameters reside in the initial metric---the model becomes more sensitive to minor parameter adjustments.
This results in an element of unpredictability with regard to the behavior of the chi-square function.

The authors of Ref.~\refcite{15} propose an useful algorithm for the holographic model calibration with the help of machine learning. As it is schematically descriped on fig. 2,3 the 2 main stages can be distinguished.

\begin{figure}[th]
\centerline{
\includegraphics[scale=0.3]{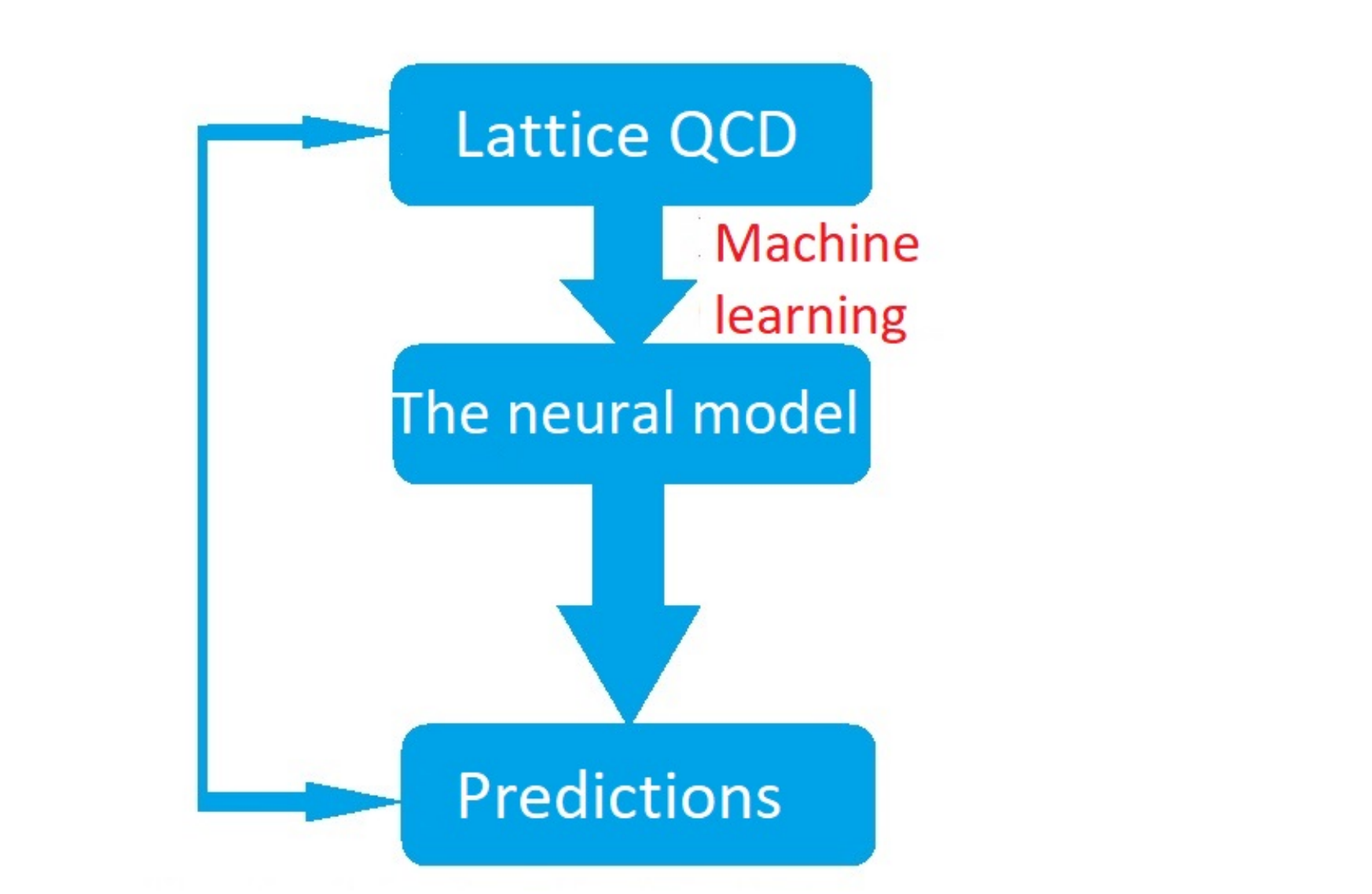}
}
\caption{The algortihm of solving the first task of calibration}
\label{f3}
\end{figure}

\begin{figure}[th]
\centerline{
\includegraphics[scale=0.3]{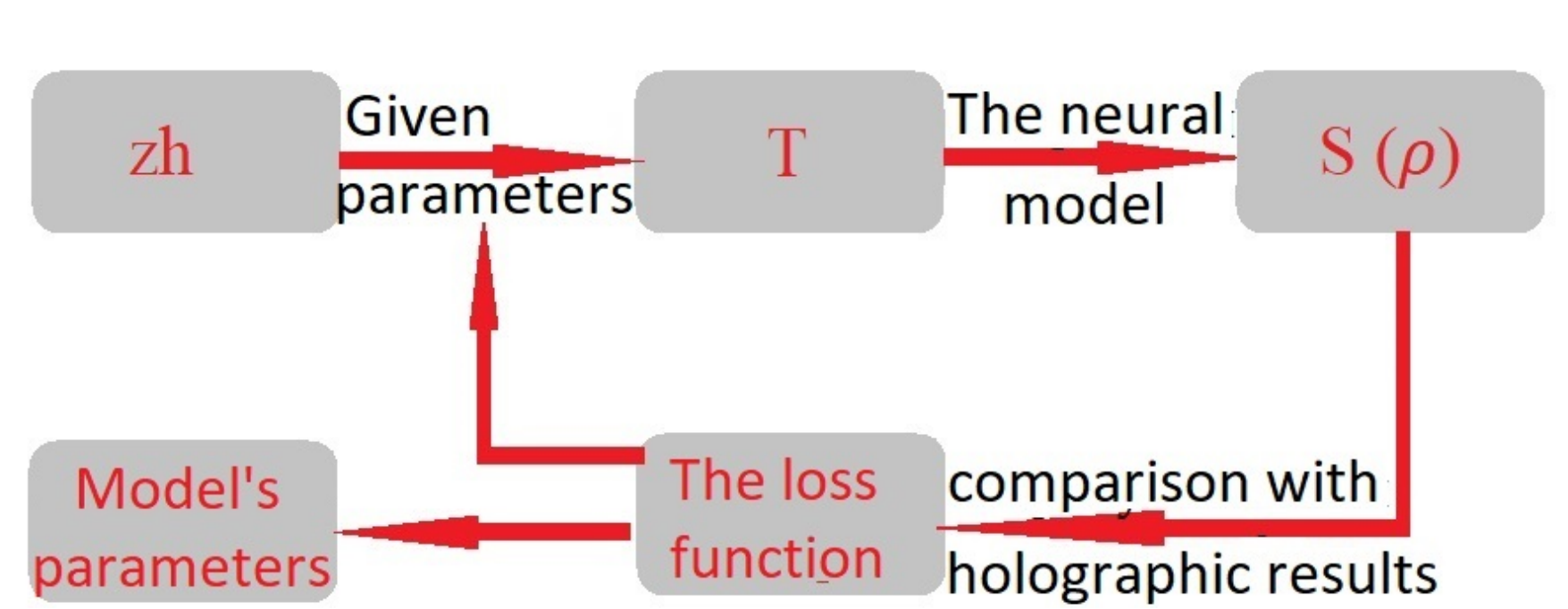}
}
\caption{The algortihm of solving the second task of calibration}
\label{f4}
\end{figure}

In the first step, a standard regression problem is solved using a neural network. For these purposes, a set of lattice data is prepared to train the model, which was used for tuning in the previous task. A sequential model consisting of four connected layers is used. Each layer uses the "ReLU" activation function. The first level contains 64 nodes, the second contains 128 nodes, and 64 nodes are allocated for the third level, all with ReLU activation. The last layer consists of a single node that corresponds to the output variable. An MSE-type loss function and the Adam algorithm are used as an optimizer, while the learning rate is set at 0.001 for effective gradient descent. The neural model, trained on the input data set, turns out to be able to make accurate predictions over the entire range of lattice data.

The second step involves solving the optimization problem using the gradient descent algorithm to find the optimal parameter values. The functions A(z), T(zh) and S(zh) are defined holographically. Then the loss function Loss(a, b, d, G) is determined, expressing the difference between the predictions of the holographic model and the values obtained from the neural network model configured in the previous step and used as a reference for parameter settings instead of a set of points from lattice calculations. The least squares method is used to minimize this difference. Some initial parameter values are determined and the ”Adam” optimizer is used for training. Since the temperature value for which the neural network trained in the first step makes predictions depends on $z_h$ in the holographic model, the input data for the optimization problem is a tensor array of gravitational radius values, which allows one to obtain an array of temperatures for each new cycle of parameter calculation.  The coefficients for the algorithm mentioned above are shown in Table 2.

\begin{table}[pt]
\tbl{Values of parameters obtained from the least squares method for alternative deforming factor\label{tab2}}
{\begin{tabular}{@{}cccccccc@{}}
\toprule
Model (alternative)       & $\nu$ & a     & b     & d      & G    & L, GeV & c     \\ \colrule
Isotropic   & 1     & 0.012 & 0.034 & -0.267 & 0.41 & 1.02   & 0.218 \\ 
Anisotropic & 4.5   & 0.361 & 0.841 & -0.249 & 0.43 & 0.7    & 0.463 \\ 
\botrule
\end{tabular}}
\end{table}

\begin{figure}[th]
\centerline{
\includegraphics[scale=0.45]{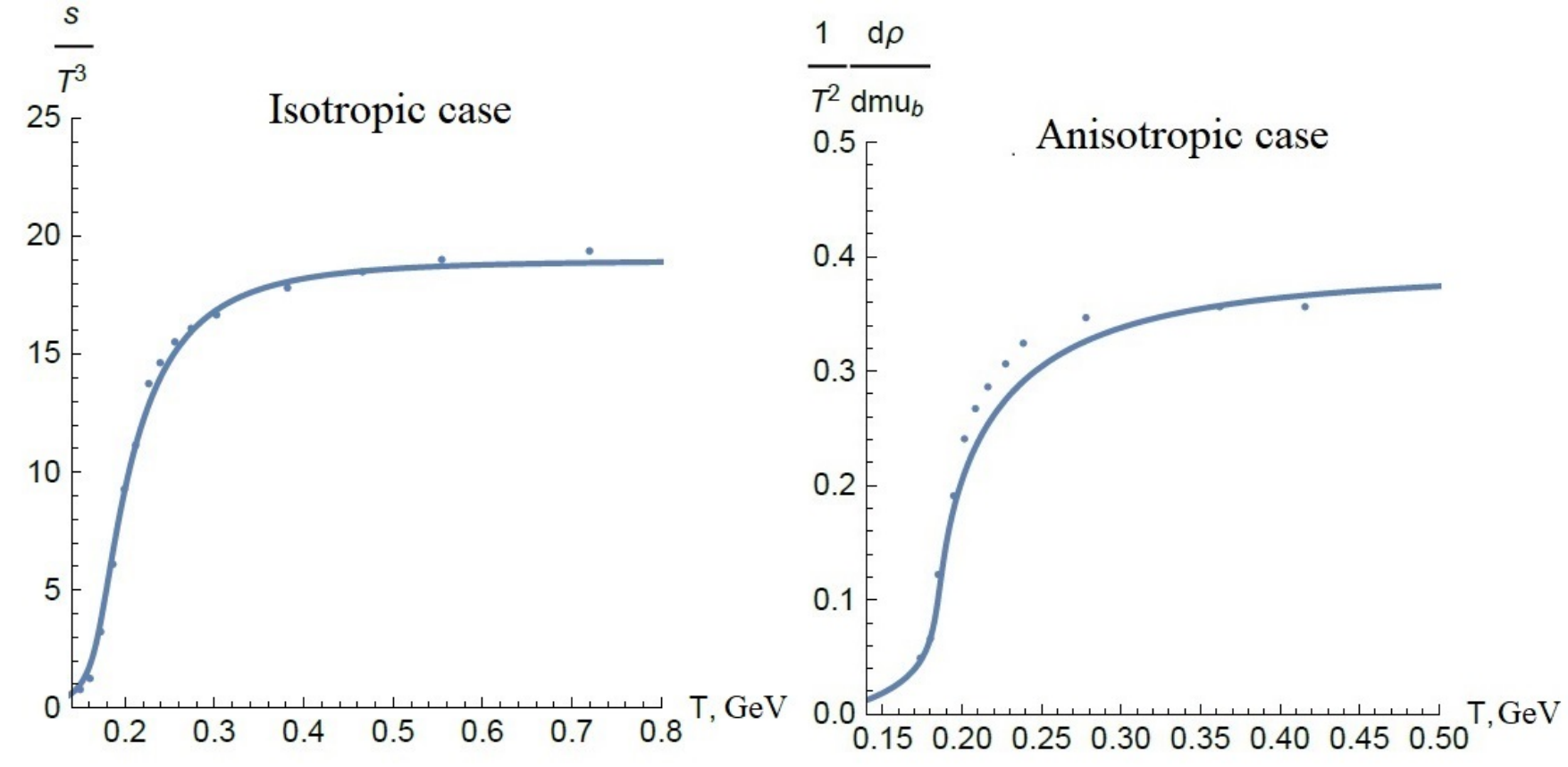}
}
\caption{Result of fitting ${s}/{T^3}$ ratio with the data from Ref.~[\citenum{12}] for isotropic (on the left) and anisotropic (on the right) models with alternative deforming factor}
\label{f5}
\end{figure}

The fit results are shown in Fig. 4.
We can see the clear improvement in the agreement of the model with Lattice QCD. The new form of deformation factor, as we believe, allows us to improve the fit qualitatively, when the tuning of free parameters with the neural network gives a better accuracy of such a calibration.

\section{The results of numerical simulation of the QGP evolution
using the holographic equation of state}

As in our previous work, we assume a qualitative correspondence between the simulation results and the calibrated lattice equations, given that the holographic model is designed to provide a complete description of the QCD phase diagram, including phase transitions.

As mentioned earlier, the complete application of the EoS occurs only in combination with the equations of relativistic hydrodynamics, which requires the use of special software packages due to the complexity of the system being solved. In the framework of this work, the relativistic hydrodynamics package MUSIC~\cite{16} is used, based on the MUSCL-type algorithm. The tables for the holographic equations of state are implemented in full compliance with the built-in lattice ones, which allows using standard linear interpolation methods for MUSIC to read the data.

It is also important that the MUSIC package is part of a larger software framework designed for step-by-step physical modeling of quark-gluon plasma evolution---from pre-equilibrium initial conditions to hadron transport models.  For this reason, our numerical simulations employ the iEBE-MUSIC package~\cite{17}, which uses the MUSIC code as its hydrodynamic core. The following calculation scheme was used:
\begin{romanlist}[(b)]
    \item The 3D MC Glauber package \cite{18}---a three-dimensional Glauber-type Monte Carlo model---generates the initial conditions for the subsequent hydrodynamical approach.
    \item The MUSIC package evolves the system using holographic equations of state until freeze-out at a specified energy density.  
    \item  The iSS package \cite{19}---performing event-by-event particle sampling from the freeze-out hypersurface--simulates particle emission while maintaining consistency between the hypersurface and transport model. 
    \item The UrQMD transport model \cite{20} handles the final stage of the simulation, including hadron re-scatterings and short-lived particle decays, providing final hadron spectra output of the entire iEBE calculation in one of the standard UrQMD formats.
\end{romanlist}

An alternative package we use is the vHLLE-SMASH hybrid approach~\cite{21}, which uses vHLLE in its core. The numerical scheme is similar:
\begin{romanlist}[(b)]
    \item SMASH \cite{22}---  transport model in the "collider" regime generates the initial conditions for hydrodynamics.
    \item The vHLLE package \cite{23} as an alternative numerical approach for hydrodynamics.  
    \item  The Hadron Sampler \cite{24}---performing particle sampling. 
    \item The SMASH transport model generates final hadron spectra for physical analysis.
\end{romanlist}

\begin{figure}[th]
\centerline{
\includegraphics[scale=0.4]{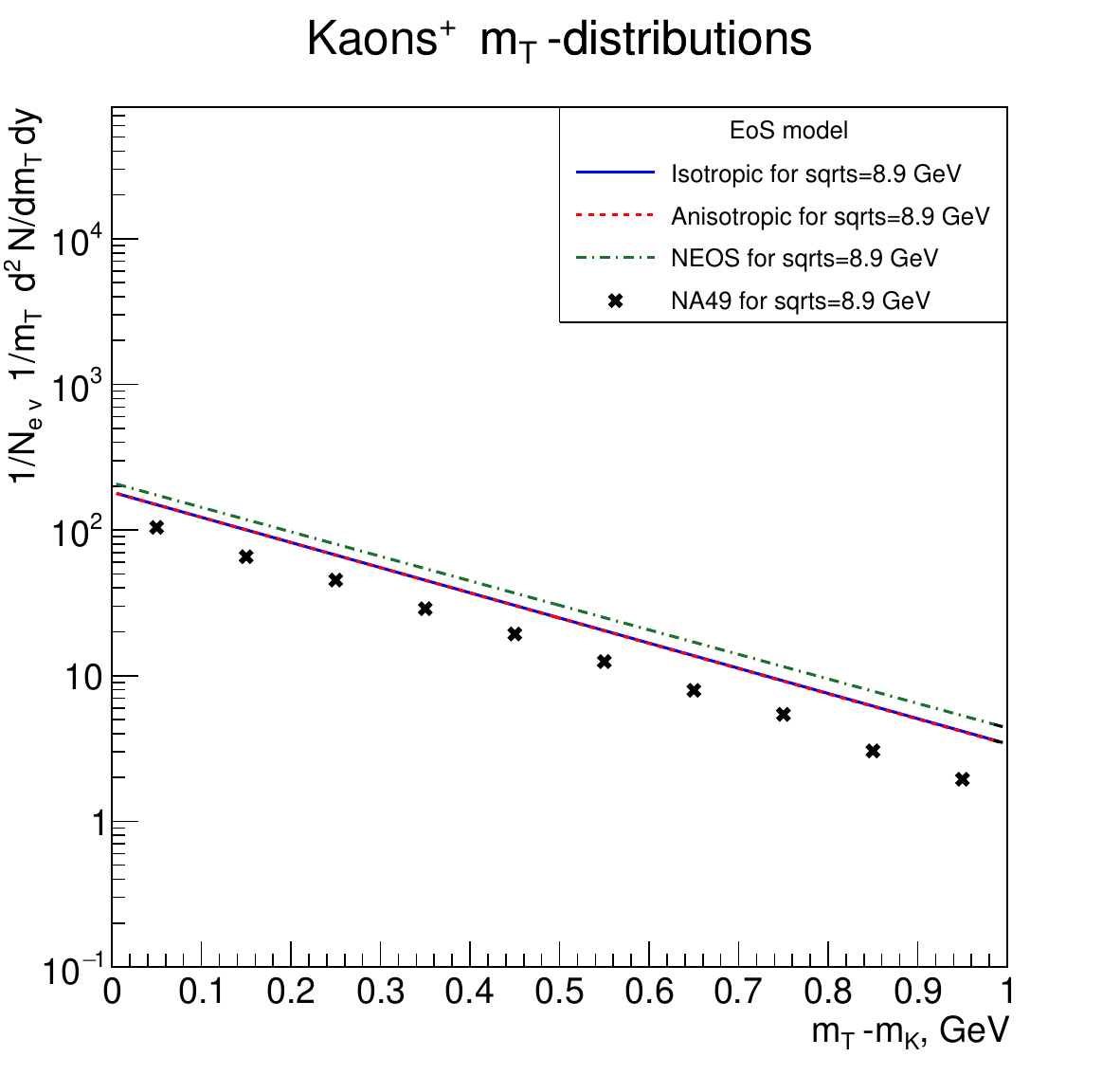}
}
\caption{Results of $m_T$ spectra for $K^{+}$ calculations within the iEBE-MUSIC package for the standard deforming factor}
\label{f7}
\end{figure}

\begin{figure}[th]
\centerline{
\includegraphics[scale=0.4]{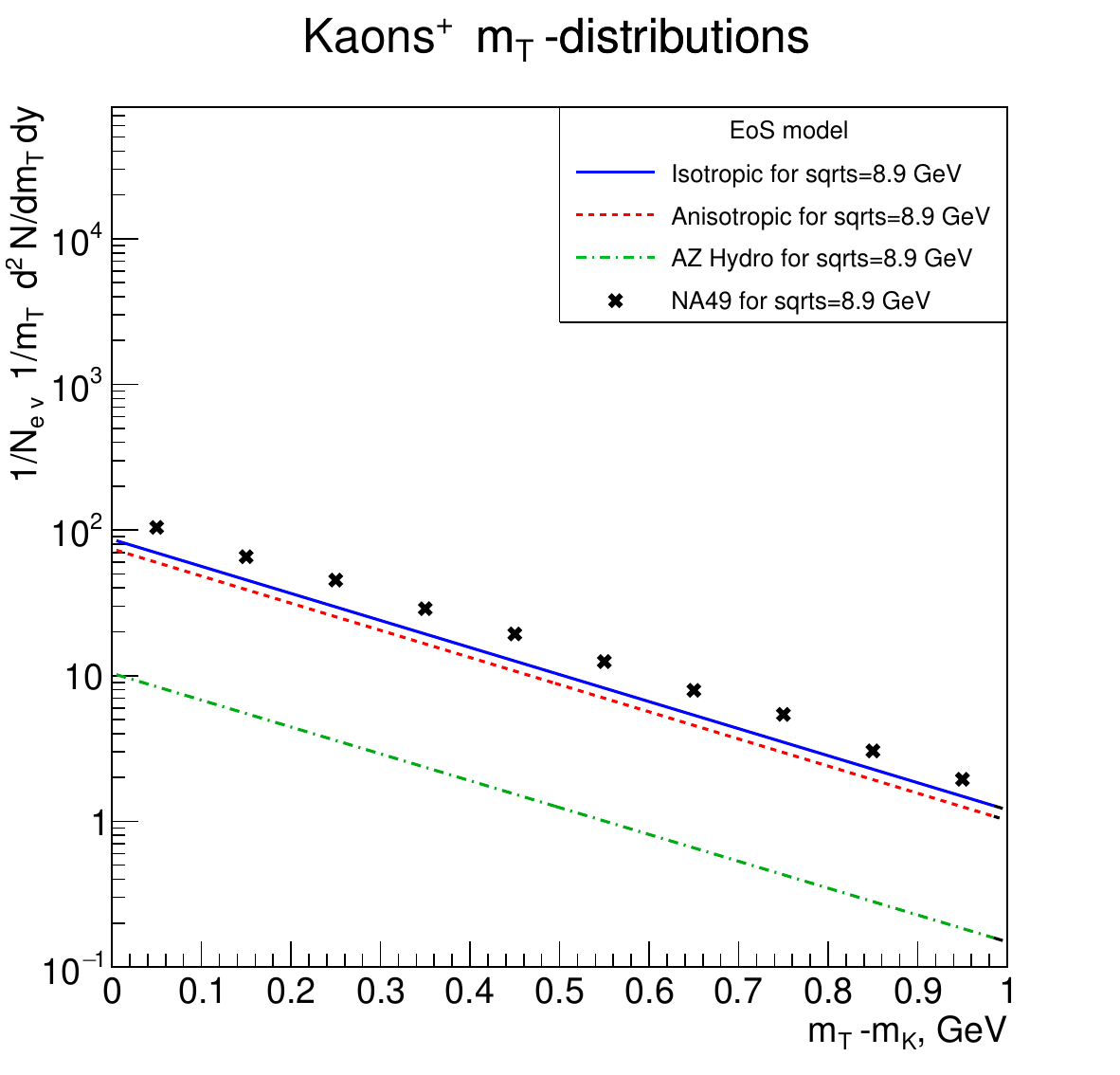}
}
\caption{Results of $m_T$ spectra for $K^{+}$ calculations within the SMASH-vHLLE package for the standard deforming factor}
\label{f8}
\end{figure}

A comparison of the corresponding predictions in the framework of standart "light quark model" deformation factor (\ref{eq4})  for the transverse mass spectra of $K^{+}$-mesons in the kinematic region of the NA49 experiment with an impact parameter of $b < 2.5$ fm is shown in Fig. 5 (iEBE-MUSIC case) and 6 (SMASH-vHLLE case). Different lines correspond to the simulation results at the energy of $\sqrt{s} = 8.9$~GeV for isotropic and anisotropic holographic equations, together with the results of the NEOS~\cite{25} equation of state, embedded in MUSIC. The latter case is taken as a reference. As a reference for the vHLLE, the AZ hydro~\cite{26} equation is used, while NEOS is internal exclusively for MUSIC.

The results for an alternative deforming factor (\ref{eq5}) are presented on fig. 7 and 8 (case of iEBE-MUSIC and SMASH-vHLLE correspondingly). Data points from fig, (5-8) from Ref.~\refcite{27} correspond to the NA49 results for the central collisions. 

\begin{figure}[th]
\centerline{
\includegraphics[scale=0.4]{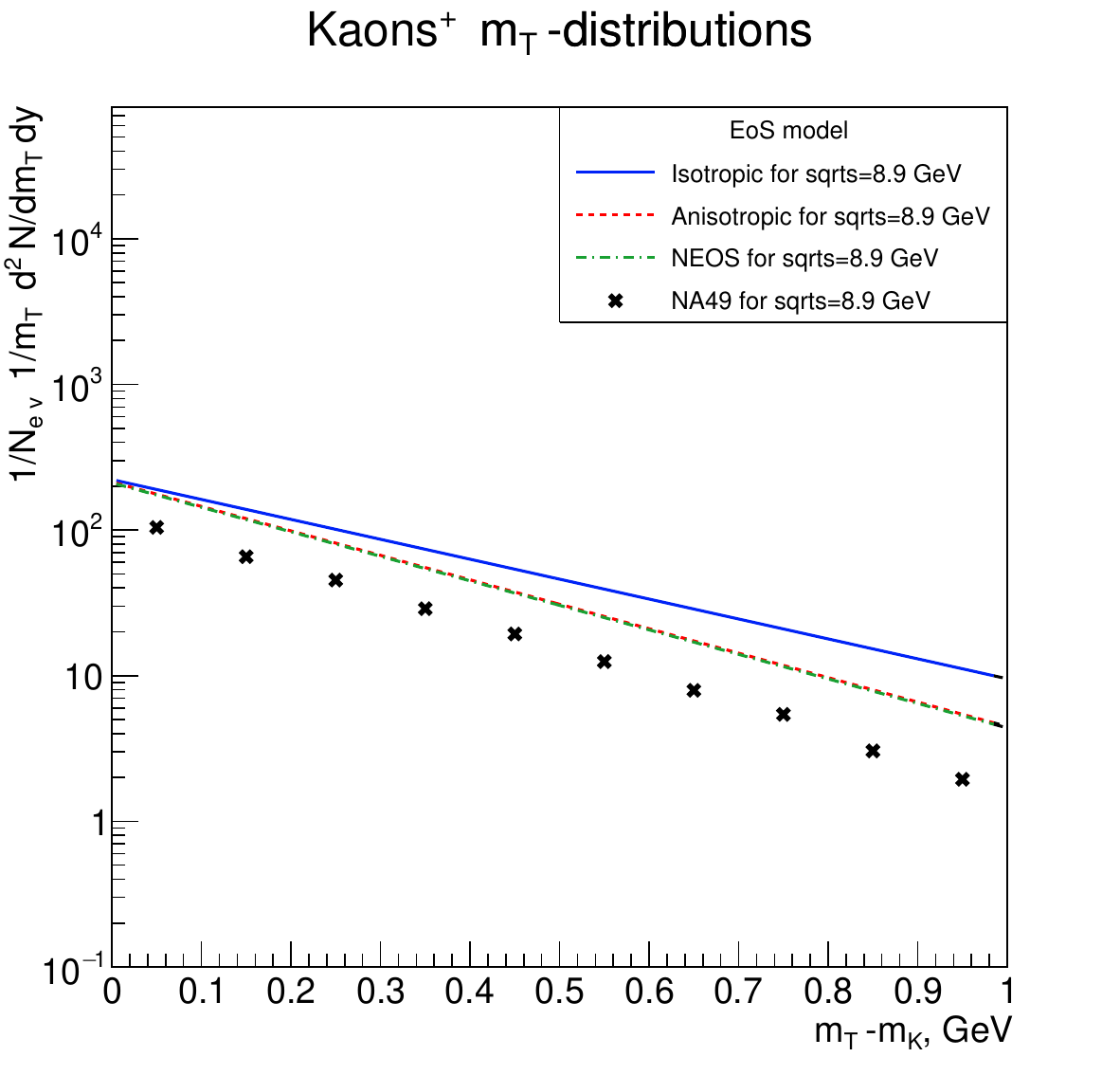}
}
\caption{Results of $m_T$ spectra for $K^{+}$ calculations within the iEBE-MUSIC package for the alternative deforming factor}
\label{f9}
\end{figure}

\begin{figure}[th]
\centerline{
\includegraphics[scale=0.4]{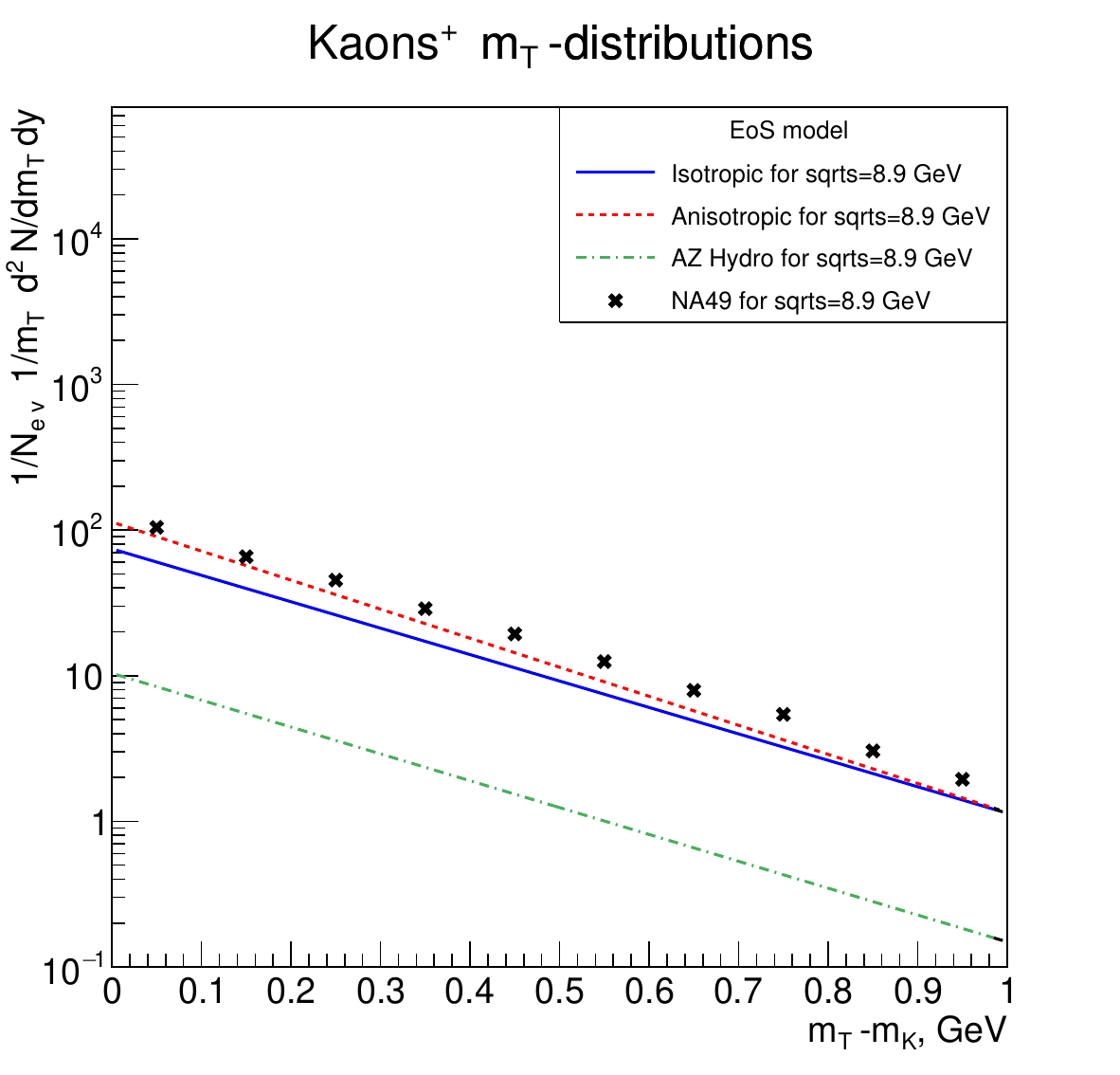}
}
\caption{Results of $m_T$ spectra for $K^{+}$ calculations within the SMASH-vHLLE package for the alternative deforming factor}
\label{f10}
\end{figure}

We observe that the transverse mass distributions in a both mentioned cases are close to each other. Taking into account the great results of calibration on fig. 5-6, we can conclude that the case of alternative deforming factor will be more useful for the next improvements of the model because of a greater agreement with the lattice at least with $\mu_b=0$. The agreement between results of numerical modeling shows that new deformation factor can be considered as a good replacement of original one for the purpose of our study,  The overall scale in hydrodynamic simulations is highly sensitive to the details of the initial conditions, so it is not considered in this paper. This does not allow us to expect a high degree of consistency of all the calculations presented with experimental data. These results demonstrate the successful calibration of both equations of state while also highlighting the need for more advanced observables that exhibit greater sensitivity to the equation of state.

\section{Conclusion and outlook}

This paper presents a method for calibrating isotropic and anisotropic holographic equations of state in the ``light quarks'' limit.  The calibration leverages lattice QCD results with physical quark masses to enable numerical simulations of matter evolution under conditions relevant to real experiments. It is shown that the two-stage tuning algorithm with machine learning described in Chapter 3 together with the use of an alternative deforming factor makes it possible to achieve significantly better agreement of the holographic model with the lattice one. The holographic EOSs are integrated into the MUSIC and vHLLE relativistic hydrodynamics package. Using the iEBE-MUSIC package and SMASH-vHLLE hybrid approach for multi-staged simulations of QGP dynamics, we find that results obtained with holographic equations of state are comparable with the lattice one. 

There is no possibility to compare our results with experimental ones because of the need for a more accurate choice of hydrodynamical parameters, which requires a special investigation. This will be done in our future works.  Furthermore, the exploration of alternative holographic models and software products can facilitate the attainment of more precise conclusions regarding the nature of the simulation outcomes that depend on the equation of state employed.

\section*{Acknowledgements}
The authors acknowledge Saint-Petersburg State University for a research project 103821868.

\section*{ORCID}

\noindent Anton Anufriev - \url{https://orcid.org/0000-0002-9155-5348}

\noindent Vladimir Kovalenko - \url{https://orcid.org/0000-0001-6012-6615}



\begin{thebibliography}{99}

\bibitem{1}

 J.\ Adams et al. (STAR Collab.), {\it Nucl. Phys. A}  {\bf 757} (2005) 102 


\bibitem{2} 

J.-Y. Ollitrault, {\it Eur. J. Phys.} {\bf  29} (2008) 275

\bibitem{3} 

Bazavov A et al. (HotQCD), {\it Phys. Rev. D} {\bf  90} (2014) 094503

\bibitem{4}  

D.\ Stephen, D. Reeb, {\it Int. J. Mod. Phys. A}, {\bf  25} (2010) 53

\bibitem{5} 

J. Maldacena, {\it Adv. Theor. Math. Phys.} {\bf  2} (1998) 231

\bibitem{6} 

J. Erlich, E. Katz, D. T. Son, M. A. Stephanov, {\it Phys. Rev. Lett.} {\bf  95} (2005) 261602 

\bibitem{7} 

M.Li, Y. Yang, P. Yang,  {\it Phys. Rev. D} {\bf  96} (2015) 066013

\bibitem{8}

I. Ya. Aref'eva, K. Rannu, P. Slepov, {\it JHEP} {\bf  06} (2021) 090

\bibitem{9} 

I. Ya. Aref'eva, A. Golubtsova, E. Gourgoulhon, {\it JHEP} {\bf  9} (2021) 142 


\bibitem{10} 

A.\ Anufriev, V. Kovalenko {\it arXiv:2504.20207} {\bf [nucl-th]} 


\bibitem{11}

 S. He,  S.-Y. Wu, Y. Yang, P.-H. Yang., {\it J. High Energ. Phys.} {\bf  04} (2013) 093 


\bibitem{12} 

M.\ Cheng et al. {\it Phys. Rev. D} {\bf  77} (2008) 014511

\bibitem{13} 

J. \ Grefa, et al.,  {\it Phys. Rev. D}, {\bf  104} (2021) 034002


\bibitem{14} 

O. DeWolfe, S. S. Gubser, C. Rosen, {\it Phys. Rev. D} {\bf  83} (2011) 086005 


\bibitem{15} 

X. \ Chen, M. Huang, {\it JHEP}, {\bf 2} (2025) 123

\bibitem{16} 

B. Schenke, S. Jeon, C. Gale, {\it Phys. Rev. C} {\bf  82} (2010) 014903 

\bibitem{17}

B. Schenke, C. Shen, P. Tribedy, {\it Phys. Rev. C} {\bf  102} (2020) 044905 

\bibitem{18}

C. Shen and B. Schenke, {\it Phys. Rev. C.} {\bf 97} (2018) 024907 

\bibitem{19}

C. Shen, Z. Qiu, H. Song, J. Bernhard, S. Bass, U. Heinz, {\it Comput. Phys. Commun.} {\bf 61} (2016) 199 

\bibitem{20}  

S. A. Bass \textit{et al.}, {\it Prog. Part. Nucl. Phys.} {\bf 41} (1998) 255 

\bibitem{21}

A. Schäfer, I. Karpenko, {\it Eur. Phys. J. A} {\bf 58} (2022) 230 


\bibitem{22}

J. Weil, V. Steinberg, J. Staudenmaier et al., {\it Phys. Rev. C} {\bf 94} (2016) 5  


\bibitem{23}

Iu. Karpenko, P. Huovinen, M. Bleicher, {\it Comput. Phys. Commun.} {\bf 3016} (2014) 185 


\bibitem{24}  

https://github.com/smash-transport/smash-hadron-sampler

\bibitem{25} 

A.\  Monnai, B. Schenke, C. Shen, {\it Phys. Rev. C} {\bf 100} (2019) 024907 

\bibitem{26} 

M. He, R. J. Fries, R. Rapp, {\it Phys. Rev. C} {\bf 85} (2012) 044911 

\bibitem{27} 

S.\ Afanasiev et al. (NA49 Collab.), {\it Phys. Rev. C} {\bf 66} (2002) 054902 


\end{thebibliography}
\end{document}